\documentclass{aastex}
\newcommand{\be}{\begin{equation}}
\newcommand{\ee}{\end{equation}}

\usepackage{spr-astr-addons}
\usepackage{url}

\RequirePackage{color}

\begin{document}

\title{ Dark energy and key physical parameters \\of
clusters of galaxies}
\slugcomment{Not to appear in Nonlearned J., 45.}
\shorttitle{Dark energy in clusters of galaxies}
\shortauthors{Bisnovatyi-Kogan and Chernin}

\author{G.S.~Bisnovatyi-Kogan\altaffilmark{1}}
\affil{Space Research Institute, Russian Academy of Sciences,
Moscow, Russia}
\and
\author{A.D.~Chernin\altaffilmark{2}}
\affil{Sternberg Astronomical Institute, Moscow University,Moscow,
Russia}

\begin{abstract}
We study physics of clusters of galaxies embedded in the cosmic
dark energy background. Under the assumption that dark energy is
described by the cosmological constant, we show that the dynamical
effects of dark energy are strong in  clusters like the Virgo
cluster. Specifically, the key physical parameters of the dark
mater halos in clusters are determined by dark energy: 1) the halo
cut-off radius is practically, if not exactly, equal to the
zero-gravity radius at which the dark matter gravity is balanced
by the dark energy antigravity; 2) the halo averaged density is
equal to two densities of dark energy; 3) the halo edge (cut-off)
density is the dark energy density with a numerical factor of the
unity order slightly depending on the halo profile. The cluster
gravitational potential well in which the particles of the dark
halo (as well as galaxies and intracluster plasma) move is
strongly affected by dark energy: the maximum of the potential is
located at the zero-gravity radius of the cluster.
\end{abstract}



\section{Introduction}

It has recently been recognized that galaxies and clusters of
galaxies (as well as all the other bodies of nature) are imbedded
in the universal dark energy background discovered first by \citet{riess98} and \citet{perl99} in observations of SNe
type Ia at the global horizon-size distances $\sim 1000$ Mpc.
These and other observations and in particular the studies of the
cosmic microwave background (CMB) anisotropy \cite{sperg07}
indicate that the global dark energy density $\rho_{\Lambda} = 0.7
\times 10^{-29}$ g/cm$^3$, and dark energy contributes nearly 3/4
to the total energy content of the universe. Close value of the
cosmological constant was anticipated by \cite{ks85},
basing on the analysis of the existing upper limits for
the microwave background anisotropy. According to the simplest,
straightforward and quite likely interpretation adopted in the
standard $\Lambda$CDM cosmology, dark energy is represented by the
Einstein cosmological constant $\Lambda$ and its density
$\rho_{\Lambda} = {\frac{c^2}{8 \pi G}} \Lambda$, where $G$ is the
gravitational constant.  If this is so, dark energy is the energy
of the cosmic vacuum \cite{glin66} and it may be described
macroscopically as a perfectly uniform fluid with the equation of
state $p_{\Lambda} = - \rho_{\Lambda}$ (here $p_{\Lambda}$ is the
dark energy pressure; the speed of light $c = 1$ hereafter). It is
this standard interpretation that implies that although dark
energy betrayed it existence through its effect on the universe as
a whole, it exists everywhere in space with the same density and
pressure.

Dark energy treated as $\Lambda$-vacuum produces antigravity, and
at the present cosmic epoch, the antigravity is stronger than the
gravity of matter for the global universe considered as a whole.
May the dynamical effects of dark energy be strong on smaller
scales as well?  Local dynamical effects of dark energy were first
recognized by \cite{cher03}; the studies of the Local
Group of galaxies and the expansion outflow of dwarf galaxies
around it revealed that the antigravity may dominate over the
gravity at distance of $\simeq 1-3$ Mpc from the barycenter of the
group \citep{cher01, cher08, bar01, kar09, byrd07, teer08, teer10}.
It was also demonstrated that this is the smallest astronomical scale on which
the antigravity produced by the dark energy background may be
stronger than the matter gravity.

Further studies \citet{cher10} show that the nearest rich
cluster of galaxies, the Virgo cluster and the Virgocentric
expansion outflow around form a system which is a scale-up version
of the Local Group with its expanding environment. It proves that
the matter gravity dominates in the volume of the cluster, while
the dark energy antigravity is stronger than the matter gravity in
the Virgocentric outflow at the distances of $\simeq 10-30 $ Mpc
from the cluster center. On both scales of 1 and 10 Mpc, the key
physical parameter of the system is its "zero-gravity radius"
which is the distance (from the system center)  where the matter
gravity and the dark energy antigravity balance each other
exactly. The gravitationally bound system can exist only within
the sphere of this radius; outside the sphere the flow dynamics is
controlled mostly by the dark energy antigravity.

\cite{abd09} used optical, X-ray and weak lensing data on
33 relaxed galaxy clusters to study the dark energy signature in the
virial structure of clusters. A new treatment has been provided for
spherically collapsing system where dark energy does not cluster
together with dark matter \citep{he10}. The spherical collapse
in quintessence models with zero speed of sound was studied
\citep{crem10}. Perturbations in DE, which is not identical with a cosmological constant, had been investigated by \citep{crem07},\citep{crem09}. The static solutions for polytropic
configurations, and their dynamic stability, in presence of the
cosmological constant, have been investigated numerically by
\citet{2012AA}.

In this paper, we focus mostly on the cluster interior. Having in
mind the Virgo cluster as an archetypical example, we consider a
cluster as a gravitationally bound quasi-spherical configuration
of cold non-relativistic collisionless dark matter and baryonic
matter in the cosmological proportion. In addition, omnipresent
dark energy with the cosmological density $\rho_{\Lambda}$ is
contained in the same volume. In Sec.2, we give a brief account of
the theory relations that describe the antigravity force field
produced by dark energy in terms of the Newtonian mechanics and
show that the zero-gravity radius may serve as a natural cut-off
radius for the dark matter halo of a cluster. This suggests some
basic interconnections that involves the matter mass, the size and
the galaxy velocity dispersion of the cluster (Sec.3). In Secs.4,5
we address physical parameters of dark halos in cluster halo and
clarify how the presence of the dark energy background restricts
their values as well as the deepness of the gravitational
potential well of the cluster. A brief discussion of the results
is given in Sec.6.

\subsection{Dark energy on the cluster scale}

Dark energy is a relativistic fluid and its description is based
on General Relativity. Nevertheless it may be treated in terms of
the Newtonian mechanics, if the force field it produces is weak in
the ordinary accepted sense. The Newtonian treatment borrows from
General Relativity the major result: the effective gravitating
density of a uniform medium is given by the sum
\be \rho_{eff} = \rho + 3 p.
\label{Eq.1}
\ee

\noindent With its equation of state $p_{\Lambda} = - \rho_{\Lambda}$, dark
energy has the negative effective gravitating density:
\be \rho_{\Lambda eff} = \rho_{\Lambda} + 3 p_{\Lambda} = - 2
\rho_{\Lambda} < 0.
\label{Eq.2}
\ee

\noindent It is because of this negative value that dark energy produces
antigravity.

With this result, one may introduce  "Einstein's law of universal
antigravity" which says that two bodies imbedded in the dark
energy background undergo repulsion from each other with the force
which is proportional to the distance $r$ between them:
\be F_{E}(r) = - {\frac{4\pi G}{3}}\rho_{\Lambda eff}r^3/r^2 = +
{\frac{8\pi G}{3}}\rho_{\Lambda}r. \label{Eq.3} \ee

\noindent (This is the force for the unit mass of the body.)
Let us consider a spherical mass $M$ of non-relativistic matter
embedded in the dark energy background. A test particle at the
distance $r$ from the mass center (and out of the mass) has the
radial acceleration in the reference frame related to the mass
center:
\be F (r) = F_N (r) + F_{E} (r) = - G \frac{M}{r^2} + {\frac{8\pi
G}{3}}\rho_{\Lambda}r.
\label{Eq.4}
\ee

\noindent Note that (\ref{Eq.4}) comes directly from the
Schwazcshild-de Sitter spacetime in the weak field approximation,
see, for instance, \citet{cher06}; (\ref{Eq.4}) may also
be used for the mass interior; in this case $M = M(r)$ in
({\ref{Eq.4}), see \citet{2012AA}.

It is seen from (\ref{Eq.4}) that the total force $F$ and the acceleration
are both zero at the distance
\be
r  =  R_{\Lambda} =
[\frac{M}{{\frac{8\pi}{3}}\rho_{\Lambda}}]^{1/3}.
\label{Eq.5}
\ee

\noindent Here $R_{\Lambda}$ is the zero-gravity radius \citep{cher03, cher01, cher08}. The gravity dominates at distances $r <
R_{\Lambda}$, the antigravity is stronger than the gravity at $r
> R_{\Lambda}$. It implies that the gravitationally bound system
with the mass $M$ can exist only within the zero-gravity sphere of
the radius $R_{\Lambda}$. Clusters of galaxies are known as the
largest gravitationally bound systems. Thus, the zero-gravity
radius is an absolute upper limit for the radial size $R$ of a static
cluster:
\be R < R_{\Lambda} =
[\frac{M}{{\frac{8\pi}{3}}\rho_{\Lambda}}]^{1/3}.
\label{Eq.6}
\ee

The total mass of the Virgo cluster estimated by \citet{kar10} with the "zero-velocity" method is $M = (6.3 \pm
2.0) \times 10^{14} M_{\odot}$. This result agrees well with the
earlier virial mass of the cluster $M_{vir} = 6 \times 10^{14}
M_{\odot}$ estimated by \citet{hs82}. \citep{teer92, ek99,ek00}
 found that the real cluster
mass $M$ might be from 1 to 2 the virial mass: $M = (0.6-1.2)
\times 10^{15} M_{\odot}$. \citet{tu04} obtained the
Virgo cluster mass $M = 1.2 \times 10^{15} M_{\odot}$. Taking for
an estimate the total mass of the Virgo cluster (dark matter and
baryons) $M = (0.6-1.2) \times 10^{15} M_{\odot}$ and the
cosmological dark energy density $\rho_{\Lambda}$ (see Sec.1), one
finds the zero-gravity radius of the Virgo cluster:
\be R_{\Lambda} = (9-11) Mpc \simeq 10 \;\;\;Mpc.
\label{Eq.7}
\ee

\noindent For the richest clusters like the Coma cluster with the
masses $\simeq 10^{16} M_{\odot}$ the zero-gravity radius may have
the values about $20$ Mpc.
Recent most systematic data on velocities and distances of
galaxies in the Virgo cluster and the Virgocentric flow around it
are presented in the Hubble diagram for the distances up to 30 Mpc
from the cluster center \citet{kar10}. The
diagram reproduced (with some modifications) in Fig.1 reveals
clearly a two-component structure in the velocity-distance ($v-r$)
phase space: there is the inner area which is the cluster itself
and the outer area of the Virgocentric flow. We countered the
areas roughly with bold dashed lines in Fig.1. The positive and
negative velocities are seen in the cluster in practically equal
numbers, so the inner component is quite symmetrical relative to
the horizontal line $v = 0$. The velocities range is -2 000 to
+1700 km/s there, and the mean velocity dispersion $V \simeq 700$
km/s. The "border" zone between the components is in the distance
range $6 < R < 12$ Mpc; it is poorly populated in the diagram, and
the velocities are considerably less scattered: they are from -400
to + 600 km/s, in the zone. The Virgocentric flow with only
positive velocities is seen at distances $ > 12$ Mpc; the flow
area is roughly symmetrical relative to a beam from the coordinate
origin $V = H_{Virgo} R$, with the local (Virgo) Hubble factor
$H_{Virgo} = 58$ km/s/Mpc.

We see that the zero-gravity radius calculated for the Virgo
cluster (\ref{Eq.5},\ref{Eq.7}) is certainly within the border zone between the two areas of the Hubble diagram: $6 < R_{\Lambda} < 12$ Mpc. This
fact suggests that the zero-gravity radius calculated for the
Virgo cluster coincides approximately, if not exactly, with the
radius of the inner component of the system in Fig.1.

\begin{figure}
\centerline{\includegraphics[scale=0.4] {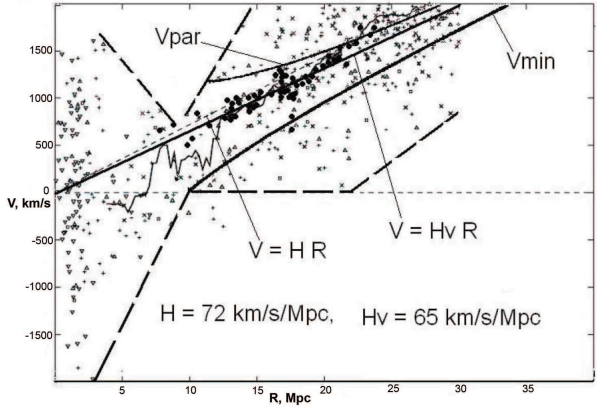} }
\caption{The Hubble diagram for 761 galaxies of the Virgo Cluster
and Virgocentric flow with Virgocentric velocities and distances
\citet{kar10}. A two-component phase structure
(countered by the bold dashed lines) is clearly seen in the
diagram: the central component is the bound quasi-stationary
cluster and the other one is the Virgocentric flow.  The matter
gravity dominates the cluster, the dark energy antigravity is
stronger than gravity in the flow. The zero-gravity radius
$R_{\Lambda} = 9-11$ Mpc is located in the less populated zone
between the components. The broken line is the running median
which used by \citet{kar10} to find the
zero-velocity radius $R_0 = 6-7$ Mpc. Two beams from the
coordinate center are the lines $V \propto R$ with the Hubble
factors $H = 72$ km/s/Mpc (the dashed one) and $H_V = 65$
km/s/Mpc. The galaxies with the most accurate distances and
velocities are marked by filled circles in the area of the
Virgocentric flow. The galaxies of this subsample follow well the
line with $H_V$ which is the median beam for them. The curve
$V_{par}$ is the flow trajectory with zero total mechanical
energy; the curve $V_{min}$ is the trajectory with the minimal
escape velocity from the potential well of the cluster. The most
accurate data occupy the area between the two curves.}
\end{figure}

\section{Cluster overall parameters}

The data of the Hubble diagram for the Virgo system (see Fig.1)
enable us to obtain another approximate empirical equality:
\be [\frac{R V^2}{GM}]_{Virgo} \simeq 1.
\label{Eq.8}
\ee

\noindent This dimensionless combination of the overall physical parameters
of the cluster resembles the traditional virial relation. However
its physical sense is  different from that of the virial
theorem, which has a form for the polytropic star with the polytropic index $n$
$$
\varepsilon_g=-\frac{n}{3}\varepsilon_g+\frac{2n}{3}\varepsilon_\Lambda,
$$

\noindent see \citet{2012AA}.

The equation (\ref{Eq.8}) does not assume any kind of
equilibrium state of the system; it does not assume either any
special relation between the kinetic and potential energies of the
system. It assumes only that the system is embedded in the dark
energy background and it is gravitationally bound.

The data on the Local Group (\citep{kar09,cher09}) show in combination with (\ref{Eq.8}) that
\be [\frac{RV^2}{GM}]_{Virgo} \simeq [\frac{RV^2}{GM}]_{LG} \simeq
1.
\label{Eq.9}
\ee

\noindent Here we use for the Local Group the following empirical
data: $R \simeq 1$ Mpc, $M \simeq 10^{12} M_{\odot}, V \simeq 70$
km/s which give the radius, the total mass and the velocity
dispersion in the Local Group, correspondingly (\citet{kar09}).

Equation (\ref{Eq.9}) indicates that there is a similarity (in the
sense of the similarity theory) between the gravitationally bound
systems of 1 Mpc scale and 10 Mpc scale. This seemingly provides
us with an evidence that such a similarity exists not only for the
Virgo System and the Local System, but for any systems of the
group and cluster scale -- at least roughly.

Assuming that the bound inner component (the cluster) of the Virgo
system has a zero-gravity radius $R_\Lambda$ (\ref{Eq.6}), we
obtain from the empirical relation (\ref{Eq.9}) that
\be V^2 \simeq\left(\frac{8\pi}{3}\right)^{1/3}
G M^{2/3}\rho_{\Lambda}^{1/3}.
\label{Eq.10}
\ee

\noindent As we see, the velocity dispersion in the gravitationally bound system
depends only on its mass and the universal dark energy density.
The relation (\ref{Eq.10}) enables one to estimate the matter mass of a
cluster, if its velocity dispersion is measured in observations:
\be M \simeq G^{-3/2} [\frac{8\pi}{3}\rho_{\Lambda}]^{-1/2} V^3
\simeq 10^{15} [\frac{V}{700 km/s}]^3 M_{\odot}.
\label{Eq.11}
\ee

\noindent From the other side, the approximate empirical relation (\ref{Eq.9})
may serve as an estimator of the local dark energy density, $\rho_{loc}$.
Indeed, if the mass of a cluster and its velocity dispersion are
independently measured, one has:
\be \rho_{loc} \simeq \frac{3}{8 \pi G^{3}} M^{-2} V^6 \simeq
\rho_{\Lambda} [\frac{M}{10^{15} M_{\odot}}]^{-2}[\frac{V}{700
km/s}]^6,
\label{Eq.12}
\ee

\noindent what indicates that the observational data on the Local System
and the Virgo System provide strong evidence in favor of the
universal value of the dark energy density which is the same on
both global and local scales.

\section{Dark matter halos}

The similarity between the Virgo cluster and the Local Group in
terms of their overall parameters (\ref{Eq.9}) does not extend to
the internal structure of the systems. As is well-known, the mass
of the Local Group is mainly contained in the two dark matter
halos of the M31 Galaxy and the Milky Way. In this sense, internal
dynamics of the group is reduced -- in the first approximation --
to the two-body problem describing the relative motion of the two
giant galaxies (see, for instance, \citet{cher09}). Contrary
to that, galaxies of the Virgo cluster move in a common dark
matter halo of the cluster as a whole, and the dark matter of the
halo is smoothly distributed over the cluster volume. Observations
and computer simulations indicate -- in acceptable agreement with
each other -- that the spherically-averaged density profiles of
the halos in various clusters are rather regular and reveal a
simple dependence on the radial distance.

In the simplest case, the halo density profile may be approximated
by the isothermal power law:
\be \rho = \rho_1 (\frac{r_1}{r})^{2},
\label{Eq.13}
\ee

\noindent where $\rho_1, r_1$ are two constants; $\rho_1 =
\rho(r_1)$. It was demonstrated (for $\Lambda$ = 0) that the
density profile of (\ref{Eq.13}) may exist in a system of particles moving
along circular orbits (\citet{bkz69}) and in
systems with almost radial orbits as well (\citet{acher75}).

According to the considerations of the section above, the
zero-gravity radius of a cluster like the Virgo cluster is roughly
(or maybe exactly) equal to the total radius of the halo. If this
is the case, we may identify $r_1$ with $R = R_\Lambda$, and
$\rho_1$ is then the dark matter density at the halo's outer edge
$\rho_{edge}$. With this, we find the total mass of the halo:
\be M = 4\pi \int^R_0 \rho r^2 dr = 4 \pi \rho_{edge}
R_{\Lambda}^{3}.
\label{Eq.14}
\ee

\noindent On the other hand, $M = \frac{8\pi}{3}
\rho_{\Lambda}R_{\Lambda}^3$; then (\ref{Eq.14}) gives:
\be \rho_{edge} = \frac{2}{3}\rho_{\Lambda}.
\label{Eq.15}
\ee

\noindent The cut-off density $\rho_1 = \rho (R_{\Lambda})$ proves
to be a constant value which does not depend on the total mass or
velocity dispersion of the isothermal halo; the density is just
the universal dark energy density with the order-of-unity
numerical factor.

It follows from (\ref{Eq.14}) that the mean halo density, $<\rho>$, is
given again by the dark energy density, but with another numerical
factor:
\be <\rho> =3\rho_{edge}= 2\rho_{\Lambda}.
\label{Eq.16}
\ee

\noindent The last relation is obviously valid for any halo's
profile, not only for the isothermal one.

Another example is a halo with the "pseudo-iso\-ther\-mal" profile:
\be \rho (r) = \frac{\rho_c}{1 + (\frac{r}{r_c})^2},
\label{Eq.17}
\ee

\noindent where $\rho_c$ is the central core density, $\rho_c =
\rho (r_c)$; and $r_c$ is the core radius.
The total mass of the halo
\be M = 4\pi \rho_c r_c^3 (\frac{R}{r_c} - \arctan\frac{R}{r_c}).
\label{Eq.18}
\ee

\noindent On the other hand, we have for the total mass a relation $M =
\frac{8\pi}{3}\rho_{\Lambda} R_{\Lambda}^3$. Then (\ref{Eq.18}) gives:
\be \rho_c = \frac{2}{3} \rho_{\Lambda} \frac{\alpha^3}{\alpha -
\arctan \alpha},
\label{Eq.19}
\ee

\noindent where $\alpha = \frac{R_{\Lambda}}{r_c}$. Here the
cut-off density
\be \rho_{edge} = \frac{\rho_c}{1 + \alpha^2} = \frac{2}{3}
\frac{\rho_{\Lambda} \alpha^3}{(\alpha - \arctan \alpha)(1 +
\alpha^2)}.
\label{Eq.20}
\ee
\noindent If the core radius is much smaller than the zero-gravity
radius, $\alpha >> 1$, (\ref{Eq.20}) gives the same result as
(\ref{Eq.16}). If, for instance, $\alpha = 3$, then $\rho_c \simeq
10 \rho_{\Lambda}$ and $\rho(R_{\Lambda}) \simeq \rho_{\Lambda}$.
If $\alpha = 10$, then $\rho_c \simeq 80 \rho_{\Lambda}$ and
$\rho(R_{\Lambda}) \simeq 0.8 \rho_{\Lambda}$.

One of possible halo profiles is suggested by computer simulations
of a large-scale structure formation -- this is the "universal"
profile introduced by \citet{nfw96}:
\be \rho (r) =  \frac{4\rho_s}{\frac{r}{r_s}(1 +
\frac{r}{r_s})^2},
\label{Eq.21}
\ee

\noindent where $\rho_s = \rho(r_s)$. The total halo mass
\be M = 16 \pi \rho_s r_s^3 [\ln (1 + \beta) - \frac{\beta}{1 +
\beta}],
\label{Eq.22}
\ee

\noindent where $\beta = \frac{R}{r_s}$. The two characteristic
densities:
\be \rho_s = \frac{1}{6} \frac{\rho_{\Lambda} \beta^3}{\ln (1 +
\beta) - \frac{\beta}{1 + \beta}},
\label{Eq.23}
\ee
\be \rho_{edge} = \frac{2}{3} \frac{\rho_{\Lambda} \beta^2}{[\ln
(1 + \beta) - \frac{\beta}{1 + \beta}] (1 + \beta)^2}.
\label{Eq.24}
\ee

\noindent If $\beta >> 1$,
\be \rho_s \simeq \frac{1}{6} \frac{\rho_{\Lambda} \beta^3}{\ln
\beta - 1},
\label{Eq.25}
\ee
\be \rho_{edge} = \frac{2}{3} \frac{\rho_{\Lambda} }{\ln \beta -
1}.
\label{Eq.26}
\ee

\noindent If $\beta =3$, we have $\rho_s \simeq 8 \rho_{\Lambda}$ and
$\rho_{edge} \simeq 0.5 \rho_{\Lambda}$. If $\beta = 10$, then
$\rho_s \simeq 20 \rho_{\Lambda}$ and $\rho_{edge} \simeq 0.2
\rho_{\Lambda}$.

Finally, the Einasto profile
(\citet{ein89}) seems to be currently most popular:
\be \rho(r) = \rho_c \exp[-(\frac{r}{r_e})^n],
\label{Eq.27}
\ee

\noindent where $\rho_c = \rho(0)$, and the constant exponent $n$
lies in between 0.1 and 0.3. Putting $r_e
=\frac{1}{N}R_{\Lambda}$, where $N \ge 1$ is a constant, and
suggesting $r_{edge}=R_\Lambda$ we find the central density:
\be \rho_c = \frac{2}{3}\frac{N^3}{I(n,N)} \rho_{\Lambda},
\label{Eq.28}
\ee

\noindent where
\be I(n,N) = \int^N_0 x^2 \exp[-x^{n}] dx.
\label{Eq.29}
\ee

\noindent Then the edge  density
\be \rho_{edge} = \rho_c \exp[-N^n].
\label{Eq.30}
\ee

\noindent The numbers $s = \frac{\rho_c}{\rho_{\Lambda}}$ and $q =
\frac{\rho_{edge}}{\rho_{\Lambda}}$ are given in Tables 1,2
 for $n = 0.1; 0.2; 0.3$ and $N = 1; 3;
10$.

\begin{table}
\caption{The ratio $s=\frac{\rho_c}{\rho_\Lambda}$ from (\ref{Eq.28}),
for different  parameters $n$ and $N$}
\begin{center}
\medskip
\begin{tabular}{ccccc}
\hline\noalign{\smallskip} {N}& {n} & {0.1} & {0.2}&
{0.3}\\
\noalign{\smallskip}
\hline\\
1 & & $5.261$ & $5.097$ & $4.946$  \\
3 & & $5.886$ & $6.413$ & $7.029$  \\
10 & & $6.758$ & $8.797$ & $12.09$ \\
\hline\\
\end{tabular}
\end{center}
\label{tab1}
\end{table}

\begin{table}
\caption{The ratio $q=\frac{\rho_{edge}}{\rho_\Lambda}$ from (\ref{Eq.30}),
for different parameters $n$ and $N$}
\begin{center}
\medskip
\begin{tabular}{ccccc}
\hline\noalign{\smallskip} {N}& {n} & {0.1} & {0.2}&
{0.3}\\
\noalign{\smallskip}
\hline\\
1 & & $1.9354$ & $1.875$ & $1.820$  \\
3 & & $1.928$ & $1.845$ & $1.750$  \\
10 & & $1.919$ & $1.803$ & $1.644$ \\
\hline\\
\end{tabular}
\end{center}
\label{tab2}
\end{table}

As we see, the antigravity produced by dark energy puts a clear
limit to the extension of dark matter halos in clusters: the halo
may exist only in the area $r \le R_{\Lambda}$ where the
antigravity is weaker than the gravity produced by non-vacuum
matter of the cluster. The dark energy density determines the mean
matter density of the halo and its edge (cut-off) density. These
are the three key physical parameters of clusters.

\subsection{Cluster potential well}

Clusters of galaxies are the largest gravitationally bound systems
in the universe. Each cluster is located in its gravitational
potential well which is the volume where galaxies, gas particles
and particles of the dark matter move along finite orbits. We will
show now that the structure of the potential well is strongly
affected by the dark energy background in which clusters are
embedded.

The gravitational potential $\Phi (r)$ inside the cluster comes
from the Poisson equation:
\be \Delta \Phi = 4 \pi G (\rho -2\rho_{\Lambda}).
\label{Eq.31}
\ee

\noindent Restricting
ourselves for simplicity by the model of the isothermal halo, we
find from (\ref{Eq.31}) together with (\ref{Eq.14}),(\ref{Eq.16}):
\be \frac{d\Phi}{dr} = - \frac{8 \pi G}{3} \rho_{\Lambda} r [1 -
(\frac{R_{\Lambda}}{r})^2], \;\;\;   0 \le r \le R_{\Lambda}.
\label{Eq.32}
\ee

\noindent In accordance with what was said above, the acceleration $
-\frac{d\Phi}{dr} = 0$ at $r = R_{\Lambda}$. The extremum
(maximum) of the potential $\Phi$ is located at the same distance
$r = R_{\Lambda}$.

The potential comes from integration of (\ref{Eq.32}):
\be \Phi(r) = - \frac{4 \pi G}{3} \rho_{\Lambda} r^2 [1  -
2(\frac{R_{\Lambda}}{r})^2 \ln \frac{r}{R_{\Lambda}}] + C.
\label{Eq.33}
\ee

\noindent The constant $C$ may be found from the boundary
condition at $r = R_{\Lambda}$:
\be \Phi(R_{\Lambda}) = - \frac{4 \pi G}{3} \rho_{\Lambda}
R_{ZG}^2 + C = - \frac{4 \pi G}{3} \rho_{\Lambda} R_{\Lambda}^2 -
\frac{GM}{R_{\Lambda}}. \label{Eq.34} \ee

\noindent It implies that $C = - \frac{GM}{R_{\Lambda}}$. Then the maximum
of the potential
\be \Phi_{max} = - {\frac{3}{2}}{\frac{GM}{R_{\Lambda}}} =
-\frac{3}{2} G (\frac{8 \pi}{3} \rho_{\Lambda})^{1/3}M^{2/3}.
\label{Eq.35}
\ee

\noindent The value of $\Phi_{max}$ depends on the cluster matter mass $M$
and the universal dark energy density. Its value is the same
for any halo profile. It gives a quantitative measure to the
deepness of the cluster potential well and determines the
characteristic isothermal velocity of the gravitationally bound
particles (galaxies, intracluster plasma particles and dark matter
particles) in the cluster:
\be V_{iso} = |\Phi_{max}|^{1/2} =
G^{1/2}(\frac{3}{2})^{1/2}(\frac{8\pi}{3} \rho_{\Lambda})^{1/6}
M^{1/3}
\label{Eq.36}
\ee
$$= 780 [\frac{M}{10^{15} M_{\odot}}]^{1/3}.
$$

\noindent As we see, this velocity is rather close to the mean velocity
dispersion, $V \simeq 700$ km/s, of the galaxies in the Virgo
cluster (Sec.3); $V_{iso} \simeq V$ also for the Coma cluster with
its matter mass $M \simeq 10^{16} M_{\odot}$ and $V \simeq 1000$
km/s.

The plasma isothermal temperature
\be T_{iso} = \frac{G m}{3k} V_{iso}^2 = \frac{m}{3k}(\frac{8
\pi}{3}\rho_{\Lambda})^{1/3} M^{2/3}
\label{Eq.37}
\ee
$$= 3 \times 10^7
[\frac{M}{10^{15} M_{\odot}}]^{2/3} K,
$$

\noindent where $k,m$ are the Boltzmann constant and the hydrogen
atom mass. The temperature of (\ref{Eq.37}) is roughly equal to the
temperature of the hot X-ray emitting plasma in clusters like the
Virgo cluster or the Coma cluster.

Identifying theoretical value $V_{iso}$ with the observed value $V$ for
typical clusters, we see that the matter mass of a cluster can be
estimated, if the velocity dispersion of its galaxies is measured:
\be M = (\frac{2}{3G})^{3/2}(\frac{8\pi}{3} \rho_{\Lambda})^{-1/2}
V_{iso}^3 = 10^{15} M_{\odot} (\frac{V}{780 km/s})^3.
\label{Eq.38}
\ee

\noindent The relation $M \propto V^3$ agrees well with the
empirical relation (\ref{Eq.11}).
In a similar way, the mass may be found, if the theoretical value of
the temperature $T_{iso}$ is identified with the measured
temperature of the intracluster plasma:
\be M = (\frac{3k}{Gm})^{3/2}
(\frac{8\pi}{3}\rho_{\Lambda})^{-1/2} T_{iso}^{3/2}
\label{Eq.39}
\ee
$$=(\frac{T}{2\times 10^7 K})^{3/2}\times 10^{15} M_{\odot}.
$$

\noindent Finally, if the matter mass of a cluster and its velocity
dispersion or its plasma temperature are measured independently,
(\ref{Eq.36})-(\ref{Eq.39}) enable one to estimate the local density of dark energy:
\be \rho_{loc} = \rho_{\Lambda} (\frac{M}{10^{15} M_{\odot}})^{-2}
(\frac{V}{780 km/s})^6,
\label{Eq.40}
\ee
\be \rho_{loc} = \rho_{\Lambda} (\frac{M}{10^{15}
M_{\odot}})^{-2}(\frac{T}{3 \times 10^7 K})^3.
\label{Eq.41}
\ee

\noindent It is seen from (\ref{Eq.35}),(\ref{Eq.36}), the empirical data on clusters like the
Virgo cluster or the Coma cluster are completely consistent with
our starting assumption that the local density of dark energy on
the scale of clusters of galaxies is the same as on the global
cosmological scales.

\section{Conclusions}

Dark energy is deeply involved in the physics which is behind the
observed structure and dynamics of clusters of galaxies. We show
here that:

1. The key physical parameter of cluster of galaxies is the
zero-gravity radius $R_{\Lambda} =
[\frac{M}{\frac{8\pi}{3}\rho_{\Lambda}}]^{1/3}$, where $M$ is the
total matter mass of the cluster and $\rho_{\Lambda}$ is the dark
energy density which is assumed to be the same everywhere in
space. A bound system must have a radius $R \le R_{\Lambda}$.

2. Observational data on the Virgo cluster suggest its radius $R$
is roughly, if not exactly, equal to system's zero-gravity radius
$R_{\Lambda}$. For the Virgo cluster $R \simeq R_{\Lambda} \simeq
10$ Mpc.

3. If this is the case, the mean density of cluster's dark matter
halo does not depend on the halo density profile and is determined
by the dark energy density only: $<\rho> = 2 \rho_{\Lambda}$. The
edge (cut-off) density of the halo depends slightly on the halo
profile: $\rho_{edge} = q \rho_{\Lambda}$, where $q = 2/3$ for the
isothermal halo and around this value for the "universal" and the
Einasto profiles.

4. Finally, the same considerations lead to new estimators of the
local density of dark energy (\ref{Eq.40}),(\ref{Eq.41}), if the matter mass and
the velocity dispersion (or the plasma temperature) are
independently measured in a cluster. The available observational
data show that the local density is near (or exactly equal to) the
global value $\rho_{\Lambda}$. This is a new argument in favor of
the interpretation of dark energy in terms of Einstein's
cosmological constant.

  Note, that basing on the above mentioned observational,
we would not be able to distinguish between the  cosmological
cosmological constant, quintessence field, or phantom field, because
rather narrow  interval of the constant $w$, $P=-w\varepsilon$,
obtained by observations of SN Ia and CMB fluctuations, $w=1.08 \pm
0.12$ \citep{sperg07}. The quantitative data about the
parameters of nearby galaxy clusters are known with a worse
precision, and cannot presently be useful for this purpose. Nevertheless, a consideration of DE perturbation in the quintessence, or phantom field, see \citet{crem07}, \citet{crem09}, and their influence on the large scale structure formation, could, in principle, to be able to give a possibility to distinguish between these cases.

\acknowledgments A.C. appreciates a partial support from the RFBR
grant 10-02-0178. The work of G.S.B.-K. was partially supported by
RFBR grants 08-02-00491 and 11-02-00602, the RAN Program 'Formation
and evolution of stars and galaxies' and Russian Federation
President Grant for Support of Leading Scientific Schools
NSh-3458.2010.2.

\end{document}